\documentclass[reprint,aip,pof,amsmath,amssymb,twocolumn]{revtex4-2}
\usepackage{float}
\usepackage{fancyhdr}
\usepackage{setspace}
\usepackage{graphicx}% Include figure files
\usepackage[svgnames]{xcolor}
\usepackage{dcolumn}% Align table columns on decimal point
\usepackage{bm}% bold math
\usepackage{times}

\usepackage{bm}
\usepackage{float}
\usepackage{graphicx}
\usepackage[normalem]{ulem}
\usepackage[svgnames]{xcolor}
\usepackage{multirow}
\usepackage{titlesec}
\usepackage{natbib}
\RequirePackage{lineno}
\usepackage{lineno}

\begin{document}

%\preprint{APS/123-QED}

\title{Magnetic freeze-out and anomalous Hall effect in ZrTe$_5$}

\author{Adrien Gourgout}
\affiliation{Laboratoire de Physique et d'Étude des Matériaux (ESPCI Paris - CNRS - Sorbonne Universit\'e), PSL Research University, 75005 Paris, France}

\author{Maxime Leroux}
\affiliation{LNCMI-EMFL, CNRS UPR3228, Univ. Grenoble Alpes, Univ. Toulouse, Univ. Toulouse 3, INSA-T,  Grenoble and Toulouse, France}

\author{Jean-Loup Smirr}
\affiliation{JEIP,  USR 3573 CNRS, Coll\`ege de France, PSL Research University, 11, place Marcelin Berthelot, 75231 Paris Cedex 05, France}

\author{Maxime Massoudzadegan}
\affiliation{LNCMI-EMFL, CNRS UPR3228, Univ. Grenoble Alpes, Univ. Toulouse, Univ. Toulouse 3, INSA-T,  Grenoble and Toulouse, France}

\author{Ricardo P. S. M. Lobo}
\affiliation{Laboratoire de Physique et d'Étude des Matériaux (ESPCI Paris - CNRS - Sorbonne Universit\'e), PSL Research University, 75005 Paris, France}

\author{David Vignolles}
\affiliation{LNCMI-EMFL, CNRS UPR3228, Univ. Grenoble Alpes, Univ. Toulouse, Univ. Toulouse 3, INSA-T,  Grenoble and Toulouse, France}

\author{Cyril Proust}
\affiliation{LNCMI-EMFL, CNRS UPR3228, Univ. Grenoble Alpes, Univ. Toulouse, Univ. Toulouse 3, INSA-T,  Grenoble and Toulouse, France}

\author{Helmuth Berger}
\affiliation{IPHYS, EPFL, CH-1015 Lausanne, Switzerland}

\author{Qiang Li}
\affiliation{
Department of Physics and Astronomy, Stony Brook University, Stony Brook, NY 11794-3800, USA}
\affiliation{Condensed Matter Physics and Materials Science Division, Brookhaven National Laboratory, Upton, New York 11973-5000, USA}

\author{Genda Gu}
\affiliation{Condensed Matter Physics and Materials Science Division, Brookhaven National Laboratory, Upton, New York 11973-5000, USA}

\author{Christopher C. Homes}
\affiliation{Condensed Matter Physics and Materials Science Division, Brookhaven National Laboratory, Upton, New York 11973-5000, USA}

\affiliation{National Synchrotron Light Source II, Brookhaven National Laboratory, Upton,
   New York 11973, USA}

\author{Ana Akrap}
\affiliation{Department of Physics, University of Fribourg, 1700 Fribourg, Switzerland}

\author{Benoît Fauqué}
\email{benoit.fauque@espci.fr}
\affiliation{JEIP,  USR 3573 CNRS, Coll\`ege de France, PSL Research University, 11, place Marcelin Berthelot, 75231 Paris Cedex 05, France}

\date{\today}

\begin{abstract}

The ultra-quantum limit is achieved when a magnetic field confines an electron gas in its lowest spin-polarised Landau level. Here we show that in this limit, electron doped ZrTe$_5$ shows a metal-insulator transition followed by a sign change of the Hall and Seebeck effects at low temperature. We attribute this transition to a magnetic freeze-out of charge carriers on the ionised impurities. The reduction of the charge carrier density gives way to an anomalous Hall response of the spin-polarised electrons. This behaviour, at odds with the usual magnetic freeze-out scenario, occurs in this Dirac metal because of its tiny Fermi energy, extremely narrow band gap and a large $g$-factor. We discuss the different possible sources (intrinsic or extrinsic) for this anomalous Hall contribution. %that can be further explored in many others low carrier Dirac materials.
\end{abstract}

\maketitle

\section{Introduction}

In the presence of a magnetic field, the electronic spectrum of a three-dimensional electron gas (3DEG) is quantized into Landau levels. When all the charge carriers are confined in the lowest Landau level---the so-called \emph{quantum limit}---the kinetic energy of electrons is quenched  in the directions transverse to the field. This favors the emergence of electronic instabilities, either driven by the electron-electron or electron-impurity interactions~\cite{celli1965,halperin1987,macdonald1987,yafet1956}. So far, the behavior of 3DEGs beyond their quantum limit has been explored in a limited number of low carrier density systems. Yet, different instabilities have been detected, such as a thermodynamic phase transition in graphite \cite{fauque2013,leboeuf2017,zhu2019,Marcenat2021}, a valley depopulation phase in bismuth \cite{Zhu2017Bi,Iwasa2019}, and a metal-insulator transition (MIT) in narrow-gap doped semi-conductors InSb \cite{shayegan1988} and InAs \cite{kaufman1970,alex2020giant}. The latter occurs when charge carriers are confined in the lowest spin-polarised Landau level -- the ultra-quantum limit. This transition is generally attributed to the magnetic freeze-out effect where electrons are frozen on ionized impurities \cite{yafet1956,Aronzon1990}. 

Lately, low-doped Dirac and Weyl materials with remarkable field-induced properties were discovered \cite{Moll2016,ramshaw2018,Liang2019,Tang2019,Gooth2019}. Of particular interest is the case of ZrTe$_5$. The entrance into its quantum limit regime is marked by quasi quantized Hall resistivity ($\rho_{xy}$) \cite{Tang2019} and thermoelectrical Hall conductivity ($\alpha_{xy}$) \cite{Zhang2020,Han2020}, followed by a higher magnetic field transition \cite{Liu2016,Tang2019}. This phase transition has initially been attributed to the formation of a charge density wave (CDW) \cite{Liu2016,Tang2019,Qin2020}. Such interpretation has been questioned because of the absence of thermodynamic evidence \cite{Galeski2021,Tian2021}, expected for a CDW transition. Furthermore, ZrTe$_5$ displays a large anomalous Hall effect (AHE), even though it is a non-magnetic material \cite{Liang2018,Sun2020,Liu2021,mutch2021,lozano2021anomalous}.

Here we report electrical, thermo-electrical and optical conductivity measurements over a large range of doping, magnetic field, and temperature in electron-doped ZrTe$_5$.  This allows us to track the Fermi surface evolution of ZrTe$_5$ and explain the nature of this phase transition, as well as its links with the observed AHE. We show that the onset of the field-induced transition can be ascribed to the magnetic freeze-out effect. In contrast with usually reported results, we show that the freeze-out regime of ZrTe$_5$ is characterized by a sign change of the Hall and thermoelectric effects, followed by a saturating Hall conductivity. Our results show that the magnetic freeze-out effect differs in this Dirac material as a consequence of the tiny band gap and large $g$-factor of ZrTe$_5$, that favor both an extrinsic and an intrinsic AHE of the spin-polarised charge carriers.

\section{Results}

\subsection{Fermi surface of ZrTe$_5$}

%\section{Field induced transition in the ultra-quantum limit of  ZrTe$_5$}

\begin{figure}[]
\begin{center}
\includegraphics[angle=0,width=8.5cm]{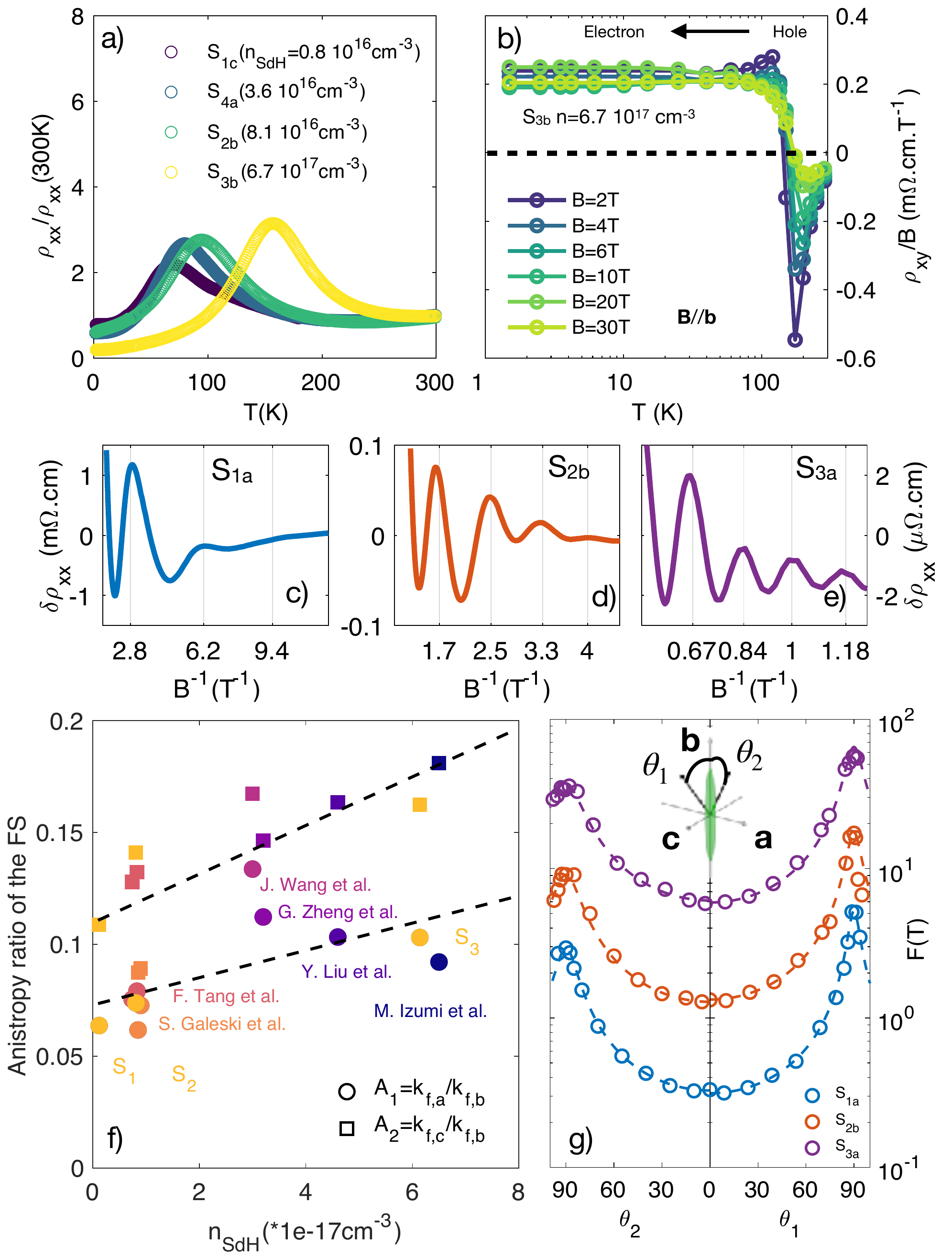}
\caption{{\bf{ Doping evolution of the Fermi surface of ZrTe$_5$: a) Temperature dependence of $\rho_{xx}$ for the four different batches studied, labelled respectively S$_{1,2,3,4}$. Samples from the same batch are labelled by distinct subscript letters. $n_{SdH}$ is the carrier density deduced from quantum oscillations (see \cite{SM}). b) $\frac{\rho_{xy}}{B}$ vs B for S$_{3b}$ (n$_{S_3}$ = 6.7 $\times$ 10$^{17}$cm$^{-3}$).  The dashed line indicates the zero value of $\rho_{xy}$.  c)-e) Shubnikov-de Haas quantum oscillations measured in the three samples S$_{1a}$, S$_{2a}$ and S$_{3b}$ at $T$ = 2 K for ${\bf{B}}\parallel{\bf{b}}$. g) Angular dependence of the frequency of quantum oscillations (F) in open circles as function of $\theta_{1,2}$, the angles between the $\bf{b}$-axis and the magnetic field rotating in the ({\bf{b}},{\bf{a}}) and ({\bf{b}},{\bf{c}}) planes,  respectively. The dotted lines are the frequency, $F$,  for an ellipsoid Fermi surface of anisotropy $A_{i}$ ($F=F_0(1+(1/A^2_i-1)cos^2(\theta))^\frac{-1}{2}$). For the two planes of rotations, $A_{i}$ is given by $\frac{k_{F,a}}{k_{F,b}}$ and $\frac{k_{F,a}}{k_{F,c}}$, labelled $A_{1}$ and $A_{2}$ respectively. Their doping evolution is shown in f) and agrees well with the literature \cite{Izumi1987,Liu2016,Zheng2016,Wang2018,Tang2019,Galeski2021}.}} }
\label{Fig1}
\end{center}
\end{figure}

Fig. \ref{Fig1}a) shows the temperature dependence of the resistivity ($\rho_{xx}$) for four batches, labelled S$_{1-4}$ respectively. Samples from the same batch are labelled by distinct subscript letters (see \cite{SM} Supplementary Note 1). At room temperature, $\rho_{xx}\approx$ 0.7 m$\Omega$.cm. With decreasing temperature, $\rho_{xx}$ peaks at a temperature around which the Hall effect ($\rho_{xy}$) changes sign, which is around 150 K for S$_{3b}$ sample (see Fig.~\ref{Fig1}b)). Both shift to lower temperature as the carrier density decreases. These effects have been tracked by laser angle-resolved photoemission spectroscopy and attributed to a temperature-induced phase transition where the Fermi energy shifts from the top of the valence band to the bottom of the conduction band as the temperature decreases \cite{Zhang2017}.

At low temperature the Fermi energy is located in the conduction band. Figs. \ref{Fig1}c)-d) show the quantum oscillations for samples from batches S$_1$, S$_{2}$ and S$_{3}$ for a magnetic field ($\bf{B}$) parallel to the $b$-axis of the orthorhombic unit cell. The angular dependence of the quantum oscillation frequency are well fitted by an anisotropic ellipsoid Fermi surface elongated along the $b$-axis, and in good agreement with previous measurements \cite{Kamm1985,Izumi1987,Galeski2021} (see Fig. \ref{Fig1}g). Our doping study reveals that the ellipsoid anisotropy increases as the system is less doped, see Fig. \ref{Fig1}f). In our lowest doped samples the ratio of the Fermi momentum (k$_F$) along the $a$ and $b$-axis reach 0.06 implying a mass anisotropy ratio of $\frac{m^{*}_{b}}{m^{*}_{a}}\simeq$ 250, where  $m^{*}_{a,b}$ are the band mass along the $a$ and $b$ axis. This large mass anisotropy ratio is comparable to the one of Dirac electrons of bismuth \cite{Zhu2011}. This Fermi surface mapping allows us to accurately determine the Fermi sea carrier densities, n$_{SdH}$, which agree well with n$_H$ (see \cite{SM} Supplemntary Note 2). Remarkably S$_1$ samples have a Hall mobility, $\mu_H$, as large as $9.7\times 10^5$ V$\cdot$cm$^{-2}\cdot s^{-1}$ and the last quantum oscillation occurs  at a small field of B$_{QL}(S_1)$ = 0.3 T for $\bf{B}\parallel{\bf{b}}$. Given the large g-factor, $g^{*}$ $\approx$ 20-30 \cite{Liu2016,Wang2018}, this last oscillation corresponds to the depopulation of the $(0,+)$ Landau level. Above it the highly mobile electrons are all confined into the lowest spin-polarised $(0,-)$ Landau level.

\subsection{Field induced transition in the ultra-quantum limit of ZrTe$_5$}

\begin{figure*}
%\begin{center}
\centering
\makebox{\includegraphics[width=1.0\textwidth]{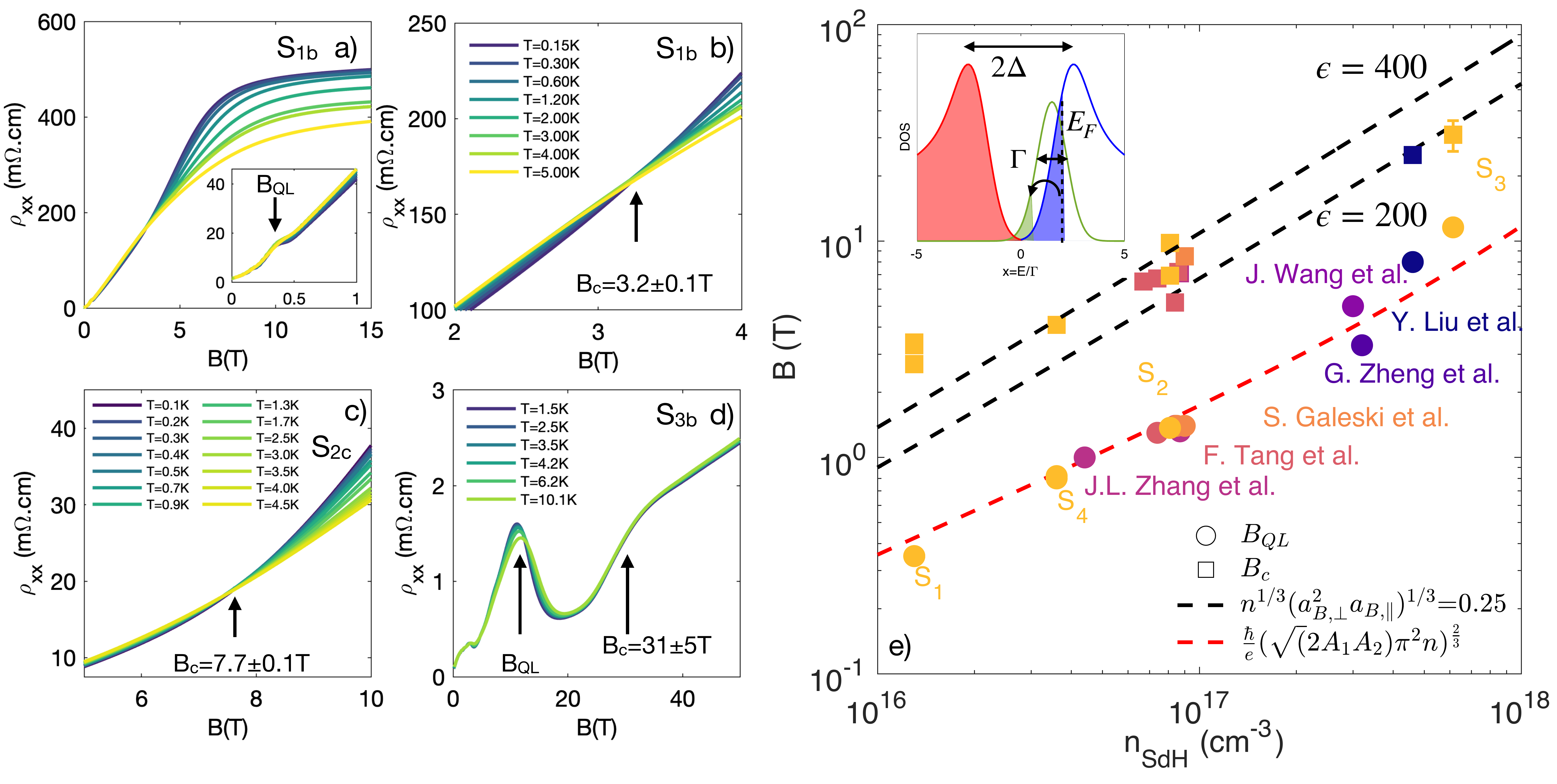}}
\caption{{\bf{Doping evolution of the transition detected in the ultra-quantum regime of ZrTe$_5$ for $\bf{B}\parallel\bf{b}$ : a) $\rho_{xx}$ vs magnetic field for samples S$_{1b}$. Inset: same as a) up to 1 T. b) Same as a) with a zoom on the crossing point in $\rho_{xx}$. c), d) Same as a) for samples  S$_{2c}$ ad S$_{3b}$. At low doping the onset of the transition is marked by a crossing point. At the highest doping it evolves into a step in $\rho_{xx}$. The width of the step has been taken as the error bar of B$_c$. e) Doping evolution of the position of the last quantum oscillation, $B_{QL}$ (yellow closed circles) and the onset of the transition, $B_{c}$ (yellow closed squares) as determined in $\rho_{xx}$ which agrees well with results from the literature \cite{Zhang2019,Tang2019,Galeski2021, Zheng2016,Wang2018,Liu2016}. The dashed red line in Fig. \ref{FigMI}e) is the value of $B_{QL}$ for an anisotropic ellipsoid (see text). The dashed black lines are the onset of the magnetic freeze-out transition according to Eq. \ref{Eq:MOT} with $\epsilon$ = 200 and 400. \textcolor{black}{Inset : sketch of the density of state of n-type ZrTe$_5$ for $B>B_c$ : the (0,-) Landau level of the conduction and valence band are plotted in blue and red while the shallow band is in green. The broadened density of states ($\Gamma$) derived from \cite{dyakonov1969}. The freez-out transition in ZrTe$_5$ occurs in the peculiar regime where $\Gamma \approx \Delta \approx E_{F}$. } }} }
\label{FigMI}
%\end{center}
\end{figure*}

Fig. \ref{FigMI} shows the field dependence of $\rho_{xx}$ beyond the ultra-quantum limit of S$_1$, S$_2$ and S$_3$ samples. In the lowest doped samples (S$_1$) $\rho_{xx}$ increases by more than two orders of magnitude and saturates above $\approx$ 7 T. This large magnetoresistance vanishes as the temperature increases (see \cite{SM}), for T$>$5 K and up to 50 T. A close inspection of the low temperature behavior reveals a light metallic phase above $B_{QL}$ (see Fig. \ref{FigMI}a) and b)) which ends at a crossing point at $B_c$ = 3.2 T above which an insulating state is observed up to 50 T. Following \cite{Tang2019} we take this crossing point as the onset of the field induced metal-insulator transition. As the carrier density increases, both the position of $B_{QL}$ and $B_{c}$ increase (see Fig. \ref{FigMI}b) and c)). At the highest doping (samples S$_3$) the amplitude of the magnetoresistance has decreased and the transition is only marked by a modest increase by a factor of two of $\rho_{xx}$ at $\simeq$ 30~T, indicating that the transition smears with increasing doping (see Fig. \ref{FigMI}d)). Fig. \ref{FigMI}e) shows the doping evolution of $B_{QL}$ and $B_{c}$ which are in good agreement with previous works \cite{Tang2019, Liu2016,Galeski2021}. \textcolor{black}{For an isotropic 3D Dirac material  $B_{QL} = \hbar/e (\sqrt(2)\pi^2n)^{2/3}$  (see i.e \cite{liang2013}) with $n=3\pi^2 k^3_{F}$. In the $\bf{B} \parallel \bf{b}$ configuration $k_F$=$\sqrt(k_{F,a}k_{F,c})$ can be evaluated from the frequency of quantum oscillations. The deduced B$_{QL}$ is shown by the red line in Fig. \ref{FigMI}e) and provides an excellent agreement with the detected $B_{QL}$. As function of the total carrier density of the ellipsoid (n$_{SdH}$) $B_{QL} = \hbar/e (\sqrt(2A_1A_2)\pi^2n_{SdH})^{2/3}$ where $A_1$ and $A_2$ are the anisotropic Fermi momentum ratios between the $a$ and $b$-axis, and between the $c$ and $b$-axis.}

%The dashed red line in Fig. \ref{FigMI}e) is the carrier density dependence of $B_{QL}$ for a 3D anisotropic Dirac pocket, given by $B_{QL}=\frac{\hbar}{e}(\sqrt{2}\pi^2\sqrt(A_{1}A_{2})n)^{\frac{2}{3}}$, where $A_1$ and $A_2$ are the anisotropic Fermi momentum ratios between the $a$ and $b$-axis, and between the $c$ and $b$-axis. Plugging the doping evolution of $A_{1,2}$ shown in Fig \ref{Fig1}f) provides an excellent agreement of the detected $B_{QL}$.

The doping evolution of $B_c$ is a clue to the nature of this transition. So far it has been attributed to the formation of a charge density wave (CDW) along the magnetic field \cite{Liu2016,Tang2019,Qin2020}. Such an instability is favored by the one-dimensional nature of the electronic spectrum along the magnetic field, which provides a suitable (2$k_F$) nesting vector in the $(0,-)$ Landau level. In this picture, predicted long ago \cite{Mermin66}, the transition is of second order and is expected to vanish as the temperature increases. The absence of temperature dependence of $B_c$ and the absence of thermodynamic signature \cite{Galeski2021,Tian2021} invite us to consider another interpretation. 
 %. Here, a$_{B0}$ is the bare Bohr radius, $\varepsilon$ the dielectric constant in units of $\varepsilon_0$, and
 
\begin{figure}[!]
\begin{center}
\includegraphics[angle=0,width=8cm]{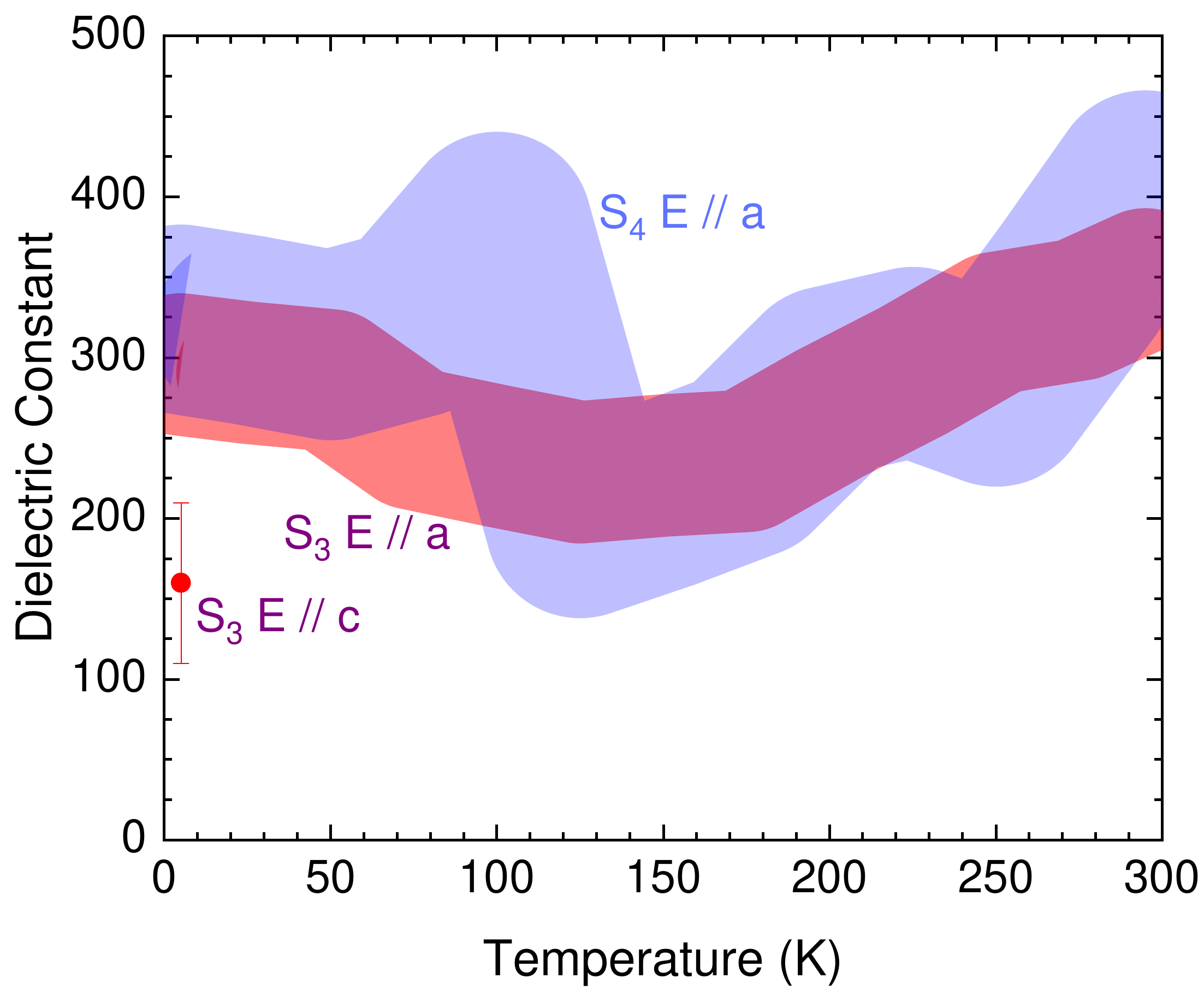}
\caption{{\bf{ Temperature dependence of the dielectric constant $\epsilon$ in units of $\epsilon_0$ for two samples from batch S$_4$ (n$_{SdH}$=3.6 $\times$ 10$^{16}$cm$^{-3}$) and S$_3$ (n$_{SdH}$=6.7 $\times$ 10$^{17}$cm$^{-3}$) for $\bf{E}\parallel \bf{a} $ and $\bf{E}\parallel \bf{c} $ (red point) of ZrTe$_5$.}} }
\label{FigEps}
\end{center}
\end{figure}

In the CDW picture, the instability is driven by the electron-electron \cite{Liu2016,Tang2019,Mermin66} or electron-phonon interaction \cite{Qin2020} and the interaction between electrons and the ionized impurities is neglected. However, in a doped semiconductor, the conduction band electrons are derived from uncompensated donors. Tellurium vacancies have been identified as the main source of impurities in ZrTe$_5$ flux grown samples \cite{Shahi2018, Salzmann2020}. According to the Mott criterion \cite{Mott1971,mott1990} a semiconductor becomes metallic when the density of its carriers, $n$, exceeds a threshold set by its effective Bohr radius, $a_{B}=4\pi\varepsilon\hbar/m^*e^2$ (where $m^{*}$ is the effective mass of the carrier, $\varepsilon$ is the dielectric constant of the semiconductor): $n^{1/3}a_{B} \simeq 0.3 $. In presence of a magnetic field the in-plane electronic wave extension shrinks with increasing magnetic field. When $B>B_{QL}$, the in-plane Bohr radius is equal to $a_{B,\perp}=2\ell_B$ with $\ell_B=\sqrt(\frac{\hbar}{eB})$ \cite{yafet1956,Shklovskii1984}. Along the magnetic field direction, the characteristic spatial extension is $a_{B,\parallel}=\frac{a_{B,z}}{\log(\gamma)}$, where $\gamma=(\frac{a_{B,c}}{l_B})^2$ with $a_{B,z}$=$\frac{\varepsilon}{m^{*}_{z}}a_{B,0}$ and $a_{B,c}$=$\frac{\varepsilon}{m^{*}_{c}}a_{B,0}$, where $m^{*}_{z,c}$ are the mass along and perpendicular to the magnetic field in units of $m_0$, and $a_{B,0}$ the bare Bohr radius. A MIT transition is thus expected to occur when the overlap between the wave functions of electrons is sufficiently decreased \cite{shayegan1988,Aronzon1990} i.e. when:
\begin{equation}
n^{1/3}(a_{B,\perp}^2a_{B,\parallel})^\frac{1}{3}\simeq0.3
\label{Eq:MOT}
\end{equation}
This MIT is thus a Mott transition assisted by the magnetic field where the metal is turned into an insulator due to the freezing of electrons on the ionized donors by the magnetic field, the so-called magnetic freeze-out effect. According to Eq.~\ref{Eq:MOT}, $n \propto B_c/\log(B_c)$ and $B_{c}$ is slightly sublinear in $n$ and evolves almost parallel to $B_{QL}$. In order to test this scenario quantitatively, one has to determine the threshold of the transition from Eq.~\ref{Eq:MOT}, which requires knowing  $\varepsilon$ and $m^{*}_{z/c}$. Temperature dependence of the quantum oscillations gives access to $m^{*}_z$ $\approx$ 2$m_0$ and $m^{*}_c$ $\approx$ 0.02$m_0$ for $\bf{B}\parallel\bf{b}$, while the optical reflectivity measurements give access to $\varepsilon$. Fig.~\ref{FigEps} shows $\varepsilon$ versus temperature for two samples of batches S$_1$ and S$_3$. $\varepsilon$ is as large as 200-400$\varepsilon_0$ in ZrTe$_5$ (see \cite{SM} Supplementary Note 5). The deduced onset from Eq.~\ref{Eq:MOT} is shown in dashed black lines in Fig.~\ref{FigMI}e) for $\varepsilon$ = 200 and 400, capturing well the doping evolution of $B_{c}$. We thus attribute the transition detected in the ultra-quantum limit of ZrTe$_5$ to the magnetic freeze-out effect. 

It is worth noticing that a large contribution to $\varepsilon$ comes from interband electronic transitions resulting in $\varepsilon_\infty >$ 100, see \cite{SM}. This result also clarifies why one can detect highly mobile carriers even down to densities as low as 10$^{13}$~cm$^{-3}$ \cite{mutch2021}. Due to the light in-plane carrier mass and large dielectric constant, one expects the threshold of the MIT at zero magnetic field to be below $\approx$ 10$^{12}$~cm$^{-3}$.

%, n$_M$=$10^{23}(\frac{m}{m_{0}\varepsilon})^{3}$~cm$^{-3}$ \cite{Mott1971,mott1990}
\section{Discussion}
\subsection{Magnetic freeze-out in ZrTe$_5$}

\begin{figure*}[]
%\begin{center}
\centering
\makebox{\includegraphics[width=1.0\textwidth]{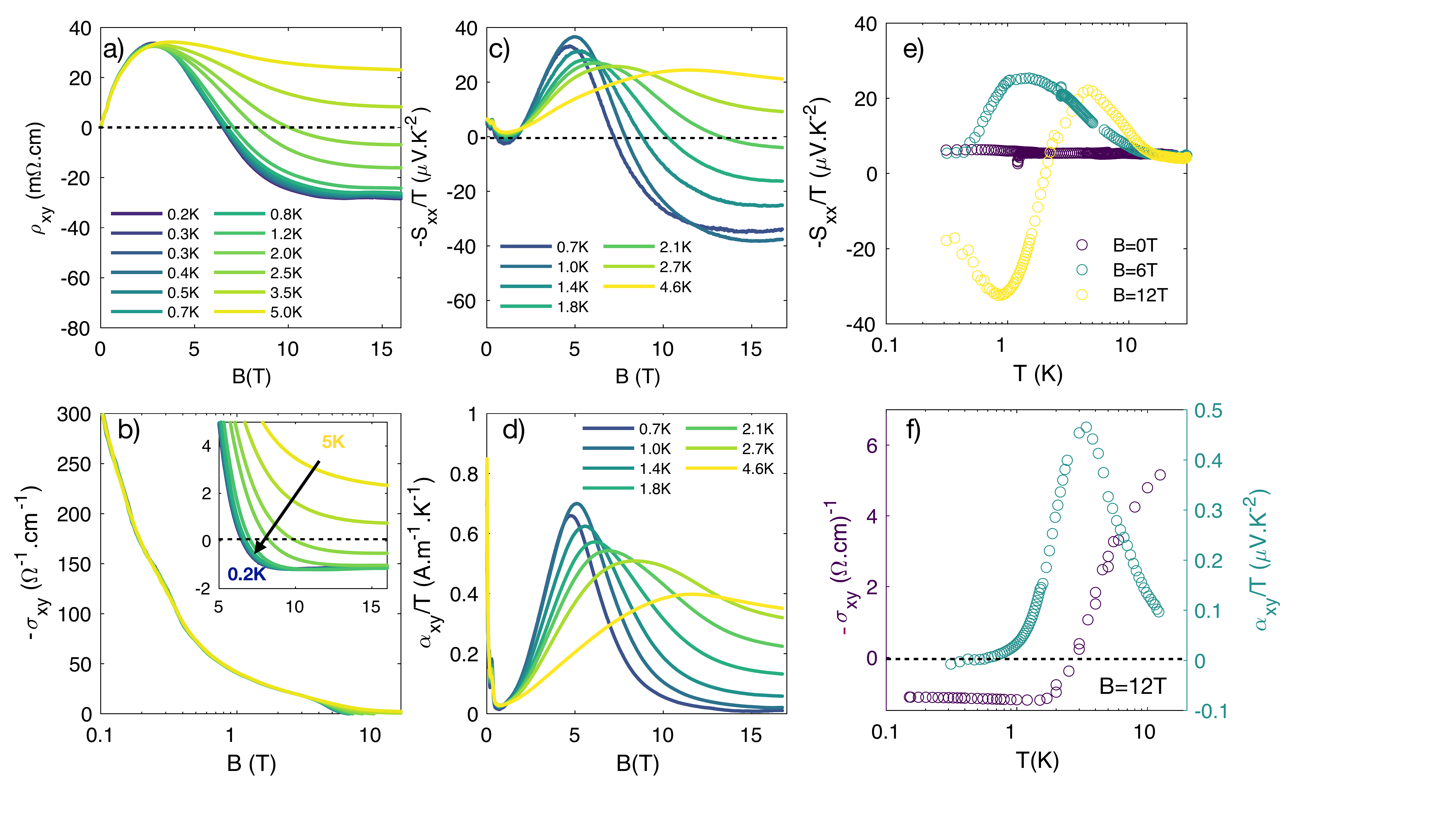}}
\caption{{\bf{ Electrical and thermoelectrical properties for S$_{1c}$ for $\bf{B}\parallel\bf{b}$: a) $\rho_{xy}$ vs $B$. b) Hall conductivity, $\sigma_{xy}$, vs $B$. \textcolor{black}{$\sigma_{xy}$=$\frac{\rho_{xy}}{\rho_{xx}*\rho_{yy}+\rho^2_{xy}}$ we assumed here that $\rho_{xx}$=$\rho_{yy}$ (see \cite{SM} Supplementary Note 4 )}. Inset of b) : zoom of $\sigma_{xy}$ from 5 to 16 T. c) Seebeck ($S_{xx}$) effect divided by the temperature vs $B$ from $T$ = 0.7 K up to 4.6 K. d)  Thermo-electrical Hall conductivity, $\alpha_{xy}$, divided by $T$ vs $B$ (see \cite{SM} Supplementary Note 4). e) Temperature dependence of $-\frac{S_{xx}}{T}$ at $B$ = 0, 6 and 12 T.  f) Temperature dependence of $\sigma_{xy}$ and $\frac{\alpha_{xy}}{T}$ at $B$ = 12 T. The sign change of $\sigma_{xy}$ is accompanied by a peak in $\frac{\alpha_{xy}}{T}$. }}}
\label{FigHall}
%\end{center}
\end{figure*}

In InSb ($n_H$= 2-5$\times 10^{15}$~cm$^{-3}$) \cite{shayegan1988}, a large drop of the carrier density comes with an activated insulating behavior. In contrast with that usual freeze-out scenario, we find in ZrTe5 a rather soft insulating behavior, where $\rho_{xx}$ saturates at the lowest temperature. Measurements of the Hall effect and thermo-electrical properties
at subkelvin temperatures shown in Fig. \ref{FigHall}a-c) reveal an unexpected field scale, thus confirming that the freeze-out regime of ZrTe$_5$ differs from the usual case. Above 7 T, $\rho_{xy}$ and the Seebeck effect ($S_{xx}$ = $\frac{-E_x}{\Delta_xt}$) change signs and saturate from 10 T up to 50 T for $\rho_{xy}$ (see \cite{SM} Supplementary Notes 3 and 4). The field induced sign changes of $\rho_{xy}$ and $S_{xx}$ are reminiscent of the sign change in temperature. 

The temperature dependence of $S_{xx}/T$ for $B$ = 0, 6 and 12 T (shown in Fig. \ref{FigHall}e) enables us to quantify the variations of charge carrier density as a function of the magnetic field. At $B$ = 0 T, $S_{xx}/T$ = $-5.5$~$\mu$V.K$^{-2}$, which is in quantitative agreement with the expected value for the diffusive response of a degenerate semiconductor: $S_{xx}/T=\frac{-\pi^2}{2}\frac{k_B}{eT_F}=-5$~$\mu$V.K$^{-2}$ for T$_F \approx $  80 K deduced from quantum oscillation measurements. At $B$ = 12 T $S_{xx}/T$ saturates, at low temperature, to $\simeq$ +20 $\mu$V.K$^{-2}$, a value which is four times larger than at zero magnetic field, pointing to a reduction of the charge carrier density by only a factor of eight. The partial freeze-out of the charge carriers is the source of the saturating $\rho_{xx}$. We now discuss the specificity of ZrTe$_5$ that leads to this peculiar freeze-out regime.

In the $k$-space, the magnetic-freeze out transition corresponds to a transfer of electrons from the lowest Landau level (0,-) to a shallow band, \textcolor{black}{see inset of Fig.\ref{FigMI}e)},  formed by the localized electrons \cite{yafet1956}. This theory does not fully apply to ZrTe$_5$ for two reasons. First, it applies to large gap systems with no potential spatial fluctuations, and ZrTe$_5$ has only a band gap of 6 meV \cite{Martino2019}, which is fifty times smaller than that of narrow gap semi-conductors such as InSb or InAs. Second, the Fermi surface of ZrTe$_5$ is highly anisotropic. The same critical field is thus reached for a carrier density that is fifty times larger in ZrTe$_5$ than in isotropic Fermi surface materials, like InSb or InAs.  The large Bohr radius and the relatively higher density of ZrTe$_5$ will therefore inevitably broaden the density of states, set by:  $\Gamma=2\sqrt{\pi}\frac{e^2}{\epsilon r_s}(N_ir^3_s)^\frac{1}{2}$ where $r_s\propto\sqrt(\frac{a_B}{n\frac{1}{3}})$ is the screening radius and N$_i$ is an estimate of the impurity concentration \cite{dyakonov1969}. Assuming that  $n \approx N_i$, we estimate $\Gamma\approx$ 6 meV in S$_1$ samples. 

In contrast with other narrow-gap semiconductors where $\Gamma << E_F << \Delta$, the magnetic freeze-out occurs in ZrTe$_5$ where $\Gamma \approx E_F \approx \Delta$. \textcolor{black}{In this limit, the shallow band of width $\Gamma$ will overlap the LLL of the conduction band, and eventually the valence band giving rise to a finite residual electron and hole charge carriers at low temperature as sketched on Fig. \ref{FigMI}e)}. As a function of doping, $\Gamma$ increases the smearing of the transition (Fig. \ref{FigMI}).  The convergence of the three energy scales $\Gamma$, $E_F$ and $\Delta$ is one source of the partial reduction of charge carrier density detected in $\rho_{xx}$, $S_{xx}$ \textcolor{black}{and of the sign change of $\rho_{xy}$}. This finite residual charge carrier should give rise to a linear Hall effect, contrasting with the saturating $\rho_{xy}$ (and $\sigma_{xy}$), which is typical of an anomalous response. We discuss this anomalous contribution in the last section.

\subsection{Anomalous Hall Effect in ZrTe$_5$}

Several studies have reported an AHE in ZrTe$_5$ \cite{Liang2018, Sun2020, Liu2021, mutch2021}. In this case, the Hall conductivity is the sum of two contributions:  $\sigma_{xy}=-\frac{ne}{B}+\sigma^{A}_{xy}$ where the first and second terms are the orbital conductivity and the anomalous Hall conductivity, respectively. At high enough magnetic field, $\sigma^{A}_{xy}$ becomes dominant, setting the amplitude and the sign of $\rho_{xy}$. So far, $\sigma^{A}_{xy}$ has been attributed to the presence of a non-zero Berry curvature---an intrinsic effect---either due to the Weyl nodes in the band structure \cite{Liang2018}, or to the spin-split massive Dirac bands with non zero Berry curvature \cite{Liu2021,mutch2021}. In the latter case, $\sigma^{A}_{xy}$ scales with the carrier density, and its amplitude is expected to be +1 ($\Omega$.cm)$^{-1}$ for $n_H$=2 $\times$ 10$^{16}$ cm$^{-3}$ \cite{Liu2021}, which is of the same order of magnitude as our results. Skew and side jump scattering are another source of AHE in non magnetic semiconductors \cite{Chazalviel1972,Nozieres1973}. Deep in the freeze-out regime of low doped InSb ($n_H$ $\approx 10^{14}$ cm$^{-3}$), a sign change of the Hall effect has been observed and attributed to skew scattering \cite{Biernat1976}. In contrast with dilute ferromagnetic alloys, where the asymmetric electron scattering is due to the spin-orbit coupling at the impurity sites, here it is caused by the spin-polarised electron scattering by ionized impurities. Its amplitude is given by $\sigma^S_{xy}$=N$_S$e$\frac{g^*\mu_B}{E_1}$, where $E_1$=$\frac{\epsilon_G(\epsilon_G+\Delta)}{2\epsilon_G+\Delta}$ with $\epsilon_G$ the band gap and $\Delta$ the spin-orbit splitting of the valence band. N$_{S}$=N$_{A}$+$n$ is the density of positively charged scattering centers with N$_A$ the density of acceptors \cite{Biernat1976}. Note that $\sigma^S_{xy}$ induces a sign change of the Hall conductivity and is only set by intrinsic parameters and by N$_{S}$. Assuming N$_S$ $\approx$ $n_H(B=0)$, and taking $g^{*}$ $\approx$ 20 \cite{Liu2016,Wang2018} and $E_1$= $\epsilon_{G}$ = 6 meV ($\epsilon_G<<\Delta$), we find that $\sigma^S_{xy}$ $\simeq$ +1 ($\Omega.cm$)$^{-1}$, which is similar to the intrinsic contribution. Remarkably, it is four orders of magnitude larger than what has been observed in low doped InSb \cite{Biernat1976}, due to the tiny gap and a (relatively) larger carrier density in ZrTe$_5$.

Therefore, the AHE contribution can induce a sign change of $\rho_{xy}$ in electron doped ZrTe$_5$. It is accompanied by a peak in $S_{xx}/T$ (see Fig. \ref{FigHall}c), $S_{xy}/T$ (see \cite{SM}) and thus in $\alpha_{xy}=\sigma_{xx}S_{xy}+\sigma_{xy}S_{xx}$ (see Fig. \ref{FigHall}d-f)). Our result shows that the thermoelectric Hall plateau \cite{Zhang2020,Han2020}, observed above 5 K, collapses at low temperature. These peaks can be understood qualitatively through the Mott relation~\cite{Behniabook} ($\frac{\alpha}{T}=-\frac{\pi^2}{3}\frac{k_B}{e}\frac{\partial \sigma(\epsilon)}{\partial \epsilon}\vert_{\epsilon=\epsilon_F}$). This is the region where $\rho_{xx}$ and $\rho_{xy}$ (and thus $\sigma_{xx}$ and $\sigma_{xy}$) change the most in field and temperature, so that $S_{xx}$ and $\alpha_{xy}$ are the largest. The increase occurs in the vicinity of $B_{c}$, causing a peak in the field dependence of $S_{xx}$ and $\alpha_{xy}$, as it happens across the freeze-out regime of InAs \cite{alex2020giant}. Whether the Mott relation can quantitatively explain the amplitude of these peaks and the sign change of $S_{xx}$ remains to be determined. \textcolor{black}{This calls to extend theoretical works \cite{skinner2018,Fu2020,Wang2021} on the electrical and thermoelectrical response to the freeze-out regime of Dirac materials such as ZrTe$_5$.}

In summary, we show that the doping evolution of the onset transition detected in the ultra-quantum limit of ZrTe$_5$ can be ascribed to the magnetic freeze-out, where electrons become bound to donors. In contrast to the usual case, the freeze-out regime of ZrTe$_5$ is marked by a modest reduction of the charge carrier density due to the convergence of three tiny energy scales in this Dirac material: the band gap, the slowly varying potential fluctuations and the Fermi energy. Deep in the freeze-out regime, the Hall conductivity changes sign and becomes anomalous with a relatively large amplitude for this low carrier density and non magnetic material. \textcolor{black}{This AHE could thus have an extrinsic origin due to skew-scattering of the spin-polarised electrons by ionized impurities}. Distinguishing and tuning both intrinsic and extrinsic contributions by varying the charge compensation or strain \cite{mutch2019} is an appealing perspective for future research. To date, the AHE of the spin-polarised electrons in the ultra-quantum limit has been detected in a limited number of cases. Many Dirac materials with small gaps and large $g$-factors remain to be studied, in particular at higher doping where the intrinsic and extrinsic AHE are both expected to be larger.

\section*{Methods}
Two sets of ZrTe$_5$ samples have been used in this study. The first ones, grown by flux method where iodine served as a transport agent for the constituents, have the lowest carrier density. The second ones, grown by Chemical Vapor Transport (CVT), have the highest density. Electrical and thermal transport  measurements have been measured using four point contacts. Contact resistance of a less 1 $\Omega$ has been achieved by an Argon etching, follow by the deposit of 10 nm Ti buffer layer and of 150 nm Pd layer. High magnetic field measurement has been done at LNCMI-Toulouse. Thermo-electrical and thermal transport measurements has been done using a standard two-thermoemeters one-heater set up similar to one used in \cite{alex2020giant}. Further experimental details can be found in \cite{SM}.

%\bibliography{biblio}
%\bibliographystyle{naturemag}

\section{Data avaliability}
All data supporting the findings of this study are available from the corresponding author B.F. upon request.

\section{Acknowledgments}

We thank K. Behnia, J-H Chu, A. Jaoui, B. Skinner and B. Yan for useful discussions. We acknowledge the support of the LNCMI-CNRS, member of the European Magnetic Field Laboratory (EMFL). This work was supported by JEIP-Coll\`{e}ge de France, by the Agence Nationale de la Recherche (ANR-18-CE92-0020-01; ANR-19-CE30-0014-04), by a grant attributed by the Ile de France regional council and from the European Research Council (ERC) under the European Union’s Horizon 2020 research and innovation program (Grant Agreement No. 636744).A.A. acknowledges funding from the  Swiss National Science Foundation through project PP00P2\_170544.
The work at Brookhaven National Laboratory was supported by the U.S. Department of Energy, Office of Basic Energy Sciences, Division of Materials Sciences and Engineering, under Contract No. DESC0012704.

\section*{Ethics declarations}
Competing interests : the authors declare no competing financial or non-financial interests.

\section*{Author contributions}
B.F and A.G conducted the electrical, thermo-electrical and thermal conductivity measurements up to $B=17$T. High-field measurements have been conducted by M.L, M.M, D.V and C.P at LNCMI-Toulouse. Optical measurements have been conducted by A.A and C.C.H and analyzed by R.L and A.A. Samples have been grown by Q.L and G. Gu. Electrical contacts on the samples have been prepared by J.L.S and B.F. B.F wrote the manuscript.

\newpage

%% Figure : 

\end{document}